\documentclass[aps,twocolumn,groupedaddress,superscriptaddress,amsmath,amssymb,pra,floatfix]{revtex4-2}

\usepackage{amsfonts,yfonts}
\usepackage{amsmath}
\usepackage{float}
\usepackage{physics}
\usepackage{mathtools}
\usepackage{bbold}
\usepackage{amsthm}

\usepackage{graphicx}
\usepackage{svg}
\usepackage{subfig}
\usepackage{braket}

\usepackage[margin=25mm]{geometry}
\usepackage{amsmath}
\usepackage{braket}
\usepackage{hyperref}

\usepackage{ mathrsfs }
\newcommand{\bqa}{\begin{eqnarray}}
\newcommand{\eqa}{\end{eqnarray}}

\usepackage{comment}
\usepackage{placeins}
\usepackage[font=small,labelfont=bf]{caption}

\newcommand{\be}{\begin{equation}}
\newcommand{\ee}{\end{equation} }

\begin{document}

\title{
Active Learning for Quantum Mechanical Measurements 
}

\author{Ruidi Zhu}
\affiliation{Department of Physics, Blackett Laboratory, Imperial College London, Prince Consort Road, SW7 2AZ, United Kingdom}
\author{Ciara Pike-Burke}
\affiliation{Department of Mathematics, Huxley Building, Imperial College London, 180 Queen's Gate, SW7 2AZ, United Kingdom}

\author{Florian Mintert}
\affiliation{Department of Physics, Blackett Laboratory, Imperial College London, Prince Consort Road, SW7 2AZ, United Kingdom}
\affiliation{Helmholtz-Zentrum Dresden-Rossendorf, Bautzner Landstraße 400, 01328 Dresden, Germany}

\begin{abstract}
The experimental evaluation of many quantum mechanical quantities requires the estimation of several directly measurable observables, such as local observables.
Due to the necessity to repeat experiments on individual quantum systems in order to estimate expectation values of observables, the question arises how many repetitions to allocate to a given directly measurable observable.
We show that an active learning scheme can help to improve such allocations, and the resultant decrease in experimental repetitions required to evaluate a quantity with the desired accuracy increases with the size of the underlying quantum mechanical system.
\end{abstract} 

\maketitle
\section{Introduction}
There is quite a discrepancy between the quantum mechanical observables that can be measured {\em in principle} and those that can be measured {\em in practice}. Restrictions to local observables or even restrictions to preferred measurement bases are common even in highly controllable experiments with synthetic quantum systems. This results in the necessity to infer quantities that can not be directly measured in terms of observables that can be measured.

While it is well understood how to express a given quantity in terms of expectation values of practically measurable observables~\cite{Yen2021Obser,James2001Qubits,Yen2020Obser,Izmaylor2019Obser}, any such decomposition implicitly assumes expectation values of all involved observables are known.
The probabilistic nature of quantum mechanical measurements, however, implies that expectation values can only be estimated with finite accuracy,
and any improvement in accuracy requires more repetitions of the same measurement.

The necessity to estimate expectation values of several observables and to increase experimental repetitions in order to improve any such estimate, opens the question of how to allocate experimental resources to the different observables.
Lacking a good basis for a different choice, each observable is typically allocated the same number of experimental repetitions~\cite{Greenaway2021Gate_uniform,Schmied2016uniform,Scott2006uniform}.

Active learning (AL)~\cite{Settles2009AL,Baldridge2004AL} is a particular type of machine learning technique aiming at optimal experimental designs. AL has been applied successfully to improve performance in many scenarios, such as speech recognition~\cite{Tur2005Language}, image retrieval~\cite{Tong2001Image}, classification tasks~\cite{Greiner2002alClassifier}, quantum information retrieval~\cite{Ding2022QuExp} and quantum state tomography~\cite{Lang2022QST}.
In contrast to the passive machine learning models~\cite{Deepmind2021Passive} (learning from randomly selected data), AL aims to minimise the number of training resources by interactively analysing the most informative samples to ensure the maximum information gain at each step. 
AL is thus ideally suited for quantum experiment design, where a crucial goal is to develop an optimal strategy in order to minimise the measurement resources.

A key ingredient in the AL scheme is to estimate the informativeness of measurement on each directly measurable observable, and hence decide which observables to query in consecutive measurements. Concentration inequalities~\cite{Boucheron2004CI,Zhao2016DeciRandom,Mnih2008EB} provide sound foundations to estimate uncertainty information of variables based on a limited amount of data, especially when this data is generated by some random variables with unknown distributions. Therefore, it is suitable to use concentration inequalities to construct query strategies based on the outcomes of quantum mechanical measurements.

In this paper, we will develop an adaptive AL scheme to decide actively which observable to measure in a repetition of an experiment in order to obtain the best-expected improvement of the estimate of a quantity that can not be directly measured.
We will show that a dynamical allocation of measurements can help to decrease the total number of repetitions required to estimate a given quantity with desired accuracy.

\section{Decisions between observables}
Most quantum mechanical quantities of interest can not be assessed in terms of a single observable.
This might be due to practical limitations, as it is the case for fidelity with respect to an entangled state or for an entanglement witness:
strictly speaking, each of those is a regular observable, but the practical restriction to measurements of single-qubit observables and correlations thereof, implies that several observables need to be measured before the expectation value of the observable of interest can be estimated~\cite{James2001Qubits,ZhangStateFidelity2021,Nielsen2000QCQI}.
This might also be due to more fundamental reasons, as it is the case with the gate fidelity~\cite{Gilchrist2005Gate,Nielsen2002Gate,Greenaway2021Gate_uniform}: since its experimental evaluation requires the implementation of a gate starting from a complete set of different initial states, estimating a gate fidelity implied performing several independent measurements even if there was not a practical restriction to local observables.

In the following, we will thus consider a general quantum mechanical quantity $Q$ whose experimental evaluation requires estimating the expectations values $\langle S_i\rangle$ of $M$ independent observables $S_i$.
Since $Q$ is a function of the expectation values $\langle S_i\rangle$, the accuracy of the estimate of $Q$ depends on the accuracy with which the expectation values $\langle S_i\rangle$ are estimated. Crucially, the accuracy of each observable does not only depend on the number of measurement repetitions, but it also depends on the actual underlying quantum state~\cite{Edwards1960Binomial}.
At any given number of repetitions of a $\sigma_z$ measurement, for example, the accuracy in the estimate of $\langle\sigma_z\rangle$ is higher for a state that is close to a $\sigma_z$ eigenstate than for a state that would yield more balanced probabilities of the two possible measurement outcomes.

Given an unknown quantum state, one can thus not find an optimal allocation of measurement repetitions to the $M$ different observables to be measured.
Only, as data is being taken, one may estimate the accuracy of the different expectation values,
and one can use this information in order to decide how to allocate measurement repetitions for subsequent experiments.

In the following, we will thus consider the situation that measurements of several observables $S_i$ that have been performed with $n_i$ repetitions each.
Based on the accumulated data, one can estimate the expectation values $\langle S_i\rangle$ and thus the value of the quantity $Q$ of interest with finite accuracy that is limited by the amount of data, {\it i.e.} the number $n_i$ of measurement repetitions.
We will derive a decision rule with AL that helps to identify the observable, the measurement of which will result in the largest available decrease of the inaccuracy in the estimate of $Q$.
With several numerical examples, we will demonstrate that taking data following this AL scheme can help to substantially decrease the number of experiments required to estimate the value of $Q$ with the desired accuracy, and that the gain grows with increasing system size.

\section{Active Learning}
\label{sec:dt}

\subsection{Concentration Inequalities}
\label{sec:ci}

The law of large numbers~\cite{Evans2004Prob} states that the expectation value $\langle S\rangle$ of a physical observable $S$ is typically approximated well by the empirical expectation value
\be
\langle S\rangle_e=\frac{1}{n}\sum_{i=1}^n s_i\ ,
\ee
where $s_i$ is the result obtained in the $i$-th repetition of the measurement of the observable $S$. The uncertainty in the estimation of $\langle S\rangle$ decreases with the number $n$ of measurement repetitions. This uncertainty can be expressed in terms of concentration inequalities, which states that the upper bound
\begin{equation} \label{eq:ci}
 |\langle S\rangle-\langle S\rangle_e|  \le \epsilon(n,\delta)
\end{equation}
on the deviation between $\langle S\rangle$ and $\langle S\rangle_e$ holds with probability $1-\delta$.

The explicit form on the upper bound $\epsilon(n,\delta)$ can depend on the underlying problem.
The Empirical Bernstein Bound \cite{Maurer2009EB,Mnih2008EB,Shivaswamy2010EB}
\begin{equation} \label{eq:EB}
    \epsilon_{B}(n,\delta) = \sqrt{\frac{2 v_e }{n}\ln\frac{2}{\delta}}+\frac{7}{3(n-1)}\ln\frac{2}{\delta}\ ,
\end{equation}
with the empirical variance
\begin{equation}
v_e=\frac{1}{n-1}\sum_{i=1}^n\left(s_i-\langle S\rangle_e\right)^2
\label{eq:empricalvariance}
\end{equation}
applies to a wide range of problems. In turn, however, it is not necessarily the best available bound for specific problems.
In the case of independent repetitions of a measurement with only two distinct outcomes (a dichotomic observable), the bound
\begin{equation} \label{eq:IEB}
    \epsilon_D(n,\delta) = \sqrt{\frac{2v}{n}  \ln \frac{1}{\delta}}
\end{equation}
with the actual variance $v$
applies~\cite{Boucheron2004CI}.
While this bound generally provides a better estimate of the accuracy of $\langle S\rangle_e$ than the Empirical Bernstein Bound, it has the disadvantage that it is not formulated in terms of the empirical variance $v_e$, but rather in terms of the actual variance
\be
v=p(1-p)({\textgoth s}_1-{\textgoth s}_2)^2\ ,
\ee
which depends on both the actual probability $p$ to obtain a distinct outcome and the values ${\textgoth s}_1$ and ${\textgoth s}_2$ that the observable $S$ can adopt ({\it i.e.} $s_i \in \{\textgoth s_1,\textgoth s_2\}$).
Since estimating the expectation value of $S$, or, equivalently, the value of the probability $p$ is the goal of the experiment,
the actual variance $v$ is indeed unknown, so that $\epsilon_D(n,\delta)$ in Eq.~\eqref{eq:IEB} is not usable in practice.

A natural remedy seems to replace the actual variance $v$ by its empirical counterpart $v_e$ as defined in Eq.~\eqref{eq:empricalvariance}.
Since, however, in cases with
close-to-certain outcomes
({\it i.e.} $p(1-p)\simeq 0$),
the empirical variance $v_e$ tends to be smaller than the actual variance $v$,
this replacement would result in an underestimate of the uncertainty of empirical expectation values of observables with low variance.
Any algorithm that is meant to decide to perform measurements of observables with uncertainty estimates would thus decide to perform too many measurements of observables with high variance and too few observables with low variance.
In order to find a decision rule that will result in close-to-optimal choices for observables to measure,
we aim at finding a rule that combines the benefits of being defined in terms of the empirical variance (as in Eq.~\eqref{eq:EB}) with the suitability to dichotomic observables (as in Eq.~\eqref{eq:IEB}). 

The heuristic ansatz 
\begin{equation} \label{eq:MB}
    \epsilon_M(n,\delta) 
    = \sqrt{\frac{2 v_{e}}{n} \ln\frac{1}{\delta}} + \frac{({\textgoth s}_1-{\textgoth s}_2)^2-4v_e}{4n} \ln\frac{2}{\delta} + \frac{1}{n}
\end{equation}
includes two additional terms as compared to Eq.~\eqref{eq:IEB}.
With their $1/n$-dependence, they become negligible in the limit $n\to\infty$.
The last term ensures that $\epsilon_M(n,\delta)$ does not vanish in the case of a few measurements ($n\gtrsim 1$).
The second term in Eq.~\eqref{eq:MB} vanishes exactly if $v_e$ adopts its maximal value, and it is the largest for vanishing empirical variance.
As such, it results in the desired modification to compensate for the misestimate of low variances.

\subsection{Uncertainty Reduction}
With the ability to estimate the uncertainty of empirical expectation values of directly measurable observables,
one can also estimate the uncertainty of the empirical estimate $Q_e$ of the composite quantity of interest.
Even though not strictly necessary, we will restrict the following discussion to functions that depend on the observables $S_i$ in a linear fashion,
since this is given for quantities like fidelity with respect to a pure state or a unitary gate.
Non-linear quantities, such as the von Neumann entropy, would require a generalisation that is feasible, but that would make the following discussion unnecessarily technical.

For any given linear function $Q=\sum_ia_i\langle S_i\rangle$, with scalar factors $a_i$, the empirical estimate of $Q$ reads  $Q_e=\sum_ia_i\langle S_i\rangle_e$, and the uncertainty of $Q_e$ can be estimated with the inequality
\be
\bigl| Q-\ Q_e\bigr|\le\sum_{i=1}^{N}|a_i|\epsilon_i\ ,
\label{eq:boundQ}
\ee
where $\epsilon_i$ is the bound on the uncertainty of $\langle S_i\rangle_e$ following Eq.~\eqref{eq:MB}.

The goal at hand is to decrease the inaccuracy of $Q_e$ through identification of the observable to measure that results in the largest possible decrease of the right-hand-side in Eq.\eqref{eq:boundQ}.
To this end, it is desirable to estimate how each of the bounds $\epsilon_i$ would change if an additional repetition of the measurement of $S_i$ was performed.

Given the dependence of the bounds in Eq.~\eqref{eq:MB} on the empirical variance, this prediction can be made only approximately.
Leaving aside situations with extremely sparse data ({\it i.e.} $n\gtrsim 1$), the change in the empirical variance following an additional measurement is expected to be negligible;
in this approximation, one can thus quantify the expected uncertainty reduction
\be
\Delta_i=\epsilon_i(n_i)-\epsilon(n_i+1)\ ,
\ee
of the estimate of {$\langle S_i\rangle_e$, where both $\epsilon_i(n_i)$ and $\epsilon(n_i+1)$ follow Eq.\eqref{eq:MB} with the empirical variance $\sigma_e$ based on $n_i$ measurements.

If all the observable $S_i$ are pairwise non-commuting, then the best available reduction in the uncertainty of $Q_e$ is achieved by measuring the observable $S_i$ that yields the largest value of $|a_i|\Delta_i$.
If there are some commuting observables within the set $\{S_i\}$, then it is essential to take into account that commuting observables can be measured in the same run of an experiment.
Instead of focusing on individual observables $S_i$, an algorithm should rather focus on groups $G_i$ of observables, such that all observables in any group do pairwise commute.
The expected uncertainty reduction of $Q_e$ upon measurement of the observables in $G_i$ is given by
\be 
w_i=\sum_{\{j|S_j\in G_i\}}|a_j|\Delta_j\ ,
\label{eq:w_i}
\ee
and the group of observables with the largest uncertainty reduction should be measured.

\subsection{Active Learning Algorithm} 
\label{sec:algo}
With the ability to identify the observables to measure that result in the largest uncertainty reduction of the empirical estimate for the quantity $Q$ of interest,
we can finally formulate the desired active learning algorithm, which is comprised of the following steps:
\begin{itemize}
\item[(i)] Since no meaningful decision can be taken without any data, it is necessary to initialise the estimation with some measurements.
As arbitrary choices should be kept to a minimum, this initialisation will be restricted to the minimal requirement to evaluate Eq.~\eqref{eq:MB}.
While a single shot is the minimum required to construct an empirical expectation value, at least two shots are required to construct an empirical variance (Eq.~\eqref{eq:empricalvariance}).
The initialisation will thus include two shots of each of the observables $S_i$ or each of the groups $G_i$.
\item[(ii)] Once there is enough data to estimate the expected uncertainty reduction $w_i$ (Eq.~\eqref{eq:w_i}), the observable $S_i$ or group $G_i$ with the largest expected reduction is selected to be measured in the next step.
The outcome of this subsequent measurement is then added to the accumulated data, and this step is repeated as long as necessary or desired.
\item[(iii)] The process of repeating step (ii) is ended if the empirical estimate $Q_e$ of $Q$ has reached the desired accuracy.
\end{itemize}

In the examples of explicit implementations of this algorithm discussed below in Sec.~\ref{sec:physical}, this process of estimating $Q_e$ will be compared with a more conventional approach,
in which step $(ii)$ is replaced by a selection of observables $S_i$ or group $G_i$ from a fixed list, such that each observable or group is measured approximately as often.

\section{Estimation of physical properties with active learning}
\label{sec:physical}

This section exemplifies the detailed process of estimating state fidelities and gate fidelities, and the dependence of the benefits of AL on the number of qubits in the underlying systems.
All of the subsequent examples are based on numerically simulated measurement outcomes, with the outcomes generated randomly following the quantum mechanical probabilities.

\subsection{State Fidelity}
\label{sec:statefidelity}
A typical example of a quantity of frequent interest is the fidelity
\be
F(\varrho,\ket{\Psi})=\bra{\Psi}{\varrho}\ket{\Psi}
\label{eq:statefidelity}
\ee
of any given state $\varrho$ with respect to a pure state $\ket{\Psi}$~\cite{James2001Qubits}.
In composite quantum systems, it can hardly ever be measured directly, but it can be cast into a weighted sum of expectation values of directly measurable observables.
For any set of mutually orthogonal observables $S_i$,
the state fidelity is of the desired form $F(\varrho,\ket{\Psi})=\sum_ia_i\bra{\Psi} S_i\ket{\Psi}$ with
\be
a_i=\frac{\tr\varrho S_i}{\tr S_i^2}\ .
\label{eq:fidelityai}
\ee
Since in most systems in the context of quantum information processing, the practically accessible observables are restricted to tensor products of Pauli matrices $\sigma_x$, $\sigma_y$, $\sigma_z$ and the identity ${\bf 1}$,
the subsequent discussion will assume this choice of observables.
Since the identity commutes with all the three Pauli matrices, and measuring an $N$-qubit observable ({\it i.e.} a tensor product of $N$ Pauli matrices, but no identity), implies also measuring all observables obtained by replacing Pauli matrices with identities without any additional effort,
this situation fits naturally into the setting of commuting observables discussed above.

In order to achieve a sound statistical comparison between fidelity estimates aided by AL and conventional methods,
the following discussion is based on state fidelity with respect to states $\ket{\Psi}$ that are randomly chosen from a distribution that is unbiased according to the Haar measure~\cite{Maziero2015RandomSampling}.
The quantum state $\varrho$ in the state fidelity Eq.~\eqref{eq:statefidelity} is chosen such that the fidelity adopts its maximal value of $1$, {\it i.e.} $\varrho=\ket{\Psi}\bra{\Psi}$, but none of the observations made in the following are specific to the case of maximal fidelity.

In particular, in the regime of few measurement repetitions, the data is strongly affected by the statistical fluctuations of measurement results.
The empirical estimates of the fidelity will thus typically vary between different realisations of the fidelity estimates.
In order to avoid substantial fluctuations in the numerical data, the subsequent discussion will therefore be based on an average over $m$ independent realisations of the same fidelity estimate for any given total number of measurement repetitions (shots) $n_T$.

With the empirical estimate $F_i(n_T)$ of the fidelity in the $i$-th realization with a given number of shots $n_T$, and the exact, theoretically constructed fidelity $F$, the standard deviation $\sigma(n_T)$ of the fidelity estimate with a given $n_T$ is defined as
\be
\sigma(n_T) = \sqrt{ \frac{1}{m-1}\sum_{i=1}^{m} \left(F-F_i(n_T)\right)^2 }\ .
\label{eq:std}
\ee
For sufficiently many repeated realisations $m$, this standard deviation is indeed independent of the statistical fluctuations that are inherent to each individual realisation,
and the subsequent examples are based on $m=10000$ realisations.

\begin{figure}[t]
\centering
\includegraphics[width=0.48\textwidth]{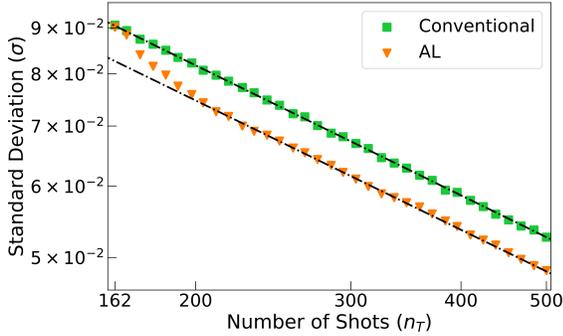}
\caption{Convergence of fidelity estimates with active learning (triangles) and with the conventionally uniform allocation of repetitions to different observables  (squares).
Both standard deviation and the number of shots are depicted on a logarithmic scale.
The lines indicate $1/\sqrt{n_T}$ convergence.}
\label{fig:convergence}
\end{figure}

Fig.~\ref{fig:convergence} depicts the convergence of such a fidelity estimate for a randomly chosen 4-qubit state with a logarithmic scale for both the number of shots $n_T$ (on the $x$-axis) and the estimated standard deviation $\sigma$ (on the $y$-axis).
Data following the estimate aided by AL is depicted with triangles, and data following the conventional estimation strategy is depicted with squares.
The black lines depict the $1/\sqrt{n_T}$ dependence that is typical for the reduction of statistical noise.
Since $2\times 3^4 =162$ shots ({\it i.e.} two measurement repetitions on each N-qubit observable) are necessary to complete the initialisation stage of the AL algorithm as described in Sec.~\ref{sec:algo}, convergence is shown only for $n_T>162$.
Initially, the convergence with the estimate aided by AL shows a faster decrease than the typical $1/\sqrt{n_T}$ dependence,
and it follows the $1/\sqrt{n_T}$-dependence only after $n_T\simeq 210$ shots. 
On the other hand, the estimate following the conventional allocation ({\it i.e.} total number of shots are evenly distributed to each
observable) follows the $1/\sqrt{n_T}$ dependence during the entire process of convergence.

The initial, faster convergence shows that the AL algorithm is indeed capable of identifying the observables to measure that best help to decrease the inaccuracy in the fidelity estimate.
Once enough data is accumulated, however, one can decide on an optimal allocation of repetitions to the different observables without accumulating more data.
In this case, the adaptive AL method can no longer outperform a strategy with a fixed, but optimised allocation, and the convergence necessarily needs to follow the $1/\sqrt{n_T}$ dependence.
Due to the initial, fast convergence, however, the approach aided by AL is expected to outperform conventional approaches also if convergence towards low variances is required, so that a larger part of the convergence is dominated by the $1/\sqrt{n_T}$ dependence.

The observation that both approaches follow the $1/\sqrt{n_T}$ dependence after sufficiently many shots (for $n_T\gtrsim 210$ in this case), is helpful to define a figure of merit for the improvement of the approach with AL over the conventional approach.
The ratio, $n_T^{(c)}/n_T^{(AL)}$, of the number of shots $n_T^{(c)}$ required to achieve a given accuracy of the fidelity with the conventional approach and the number of the shots $n_T^{(AL)}$ to achieve the same accuracy with the approach aided by AL ({\it i.e.} $\sigma(n_T^{(c)})=\sigma(n_T^{(AL)}$)), 
is independent of a desired standard deviation of the fidelity estimate as long as this standard deviation is sufficiently small so that the comparison is taken after the initial interval of fast convergence.
In the following, we will thus refer to the ratio $n_T^{(c)}/n_T^{(AL)}$ as the {\em improvement}. 

Since state fidelity can be defined for systems with various numbers of qubits, it is well suited to highlight the benefits of AL with increasing system size.
The following discussion is thus focused on the state fidelity of an $N$-qubit system with $N$ ranging from one to six.

\begin{figure}[h]
\centering
\includegraphics[width=0.48\textwidth]{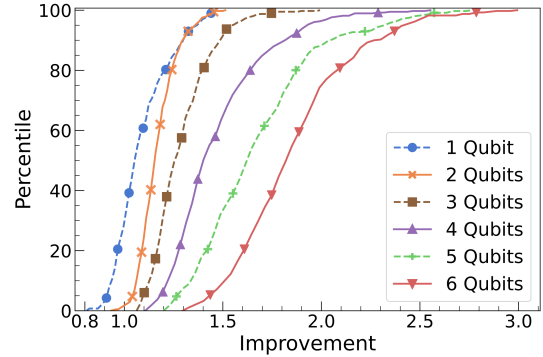}
\caption{
Cumulative distribution of improvements of state fidelity estimations with active learning obtained from statistics with $400$ random states for each system size ranging from one to six qubits.
While the improvement does depend on the underlying state, there is a clear trend of increasing improvements with a growing qubit number, due to the growing number of observables to choose from.}

\label{fig:state}
\end{figure}

Fig.~\ref{fig:state} depicts the cumulative distribution of the improvement $n_T^{(c)}/n_T^{(AL)}$ found for different system sizes based on fidelity estimates for $400$ different random states.
In the case of a single qubit (dashed line with circles), one can notice that the improvement is smaller than one in about $25\%$ of the cases.
In those cases, the conventional method yields better estimates than the method aided by AL.
The distribution of the observed improvements, however, is skewed towards higher values, and the average improvement does indeed indicate in favour of the AL method.

The only moderate benefit of AL in the estimate of single qubit fidelities can be attributed to the fact that there are only three different directly-measurable observables to choose from.
Since, however, the range of different observable settings grows exponentially in the number of qubits, one would expect that the benefits of AL become increasingly pronounced with increasing system size.
This expectation is also clearly corroborated by Fig.~\ref{fig:state}.
For three qubits and more, the improvement does always exceed the threshold value of one, and the observed improvements grow steadily with the number of qubits.
For $N=6$ qubits (solid line with downwards triangles), the improvement exceeds the value $1.8$ in half of the cases, and the improvement reaches values up to $3$;
that is, the number of measurements to be taken can be reduced by a factor of $3$ without a decrease in the accuracy of the fidelity estimate.

\subsection{Gate Fidelity}
The case of state fidelity highlights that the benefits of AL are particularly pronounced if there is a large number of measurement settings to choose from.
Since, in the case of gate fidelity, there is a choice for both the initial state and the measurement to be taken on the final state, the estimate aided by AL is potentially particularly beneficial for the estimate of gate fidelities.
This section will thus focus on the estimate of gate fidelities.
Rather than analysing statistics over randomly chosen gates, this section focuses on the two-qubit controlled-NOT (CNOT) gate and the three-qubit Toffoli gate.

The fidelity of a quantum channel $\Lambda$ with respect to a gate $U$ for N qubits~\cite{Gilchrist2005Gate} is given by
\begin{equation} \label{eq:gatefid}
    F(\Lambda,U)  = \frac{1}{2^{2N}} \sum_{i,j}
     \bra{i}U^\dagger\Lambda( |i\rangle \langle j| )\ U\ket{j}\ ,
\end{equation}
where the summation is performed over two complete sets of orthonormal state vectors.

In order to recast the definition of gate fidelity into an experimentally realisable measurement prescription, it is necessary to expand each of the operators $|i\rangle \langle j|$ in the argument of $\Lambda$ into a set of actual quantum states.
While a set of four quantum states is sufficient for a single qubit,
the following analysis is based on the five states
\begin{equation} \label{eq:basis}
    \begin{array}{l}
 \ket{\phi_0} = \ket{0}\ , \  \ket{\phi_1} = \ket{1}\  \mbox{and}\vspace{2mm} \\
 \displaystyle\ket{\phi_k} = \frac{1}{ \sqrt{2}} \left(\ket{0}+ e^{i\frac{2\pi}{3}(k-2)}\ket{1}\right)\ \mbox{for}\ k=2,3,4 \ . 
    \end{array}
\end{equation}

\begin{figure*}[t]
\centering
\subfloat[CNOT Gate]{
\includegraphics[width=0.48\textwidth]{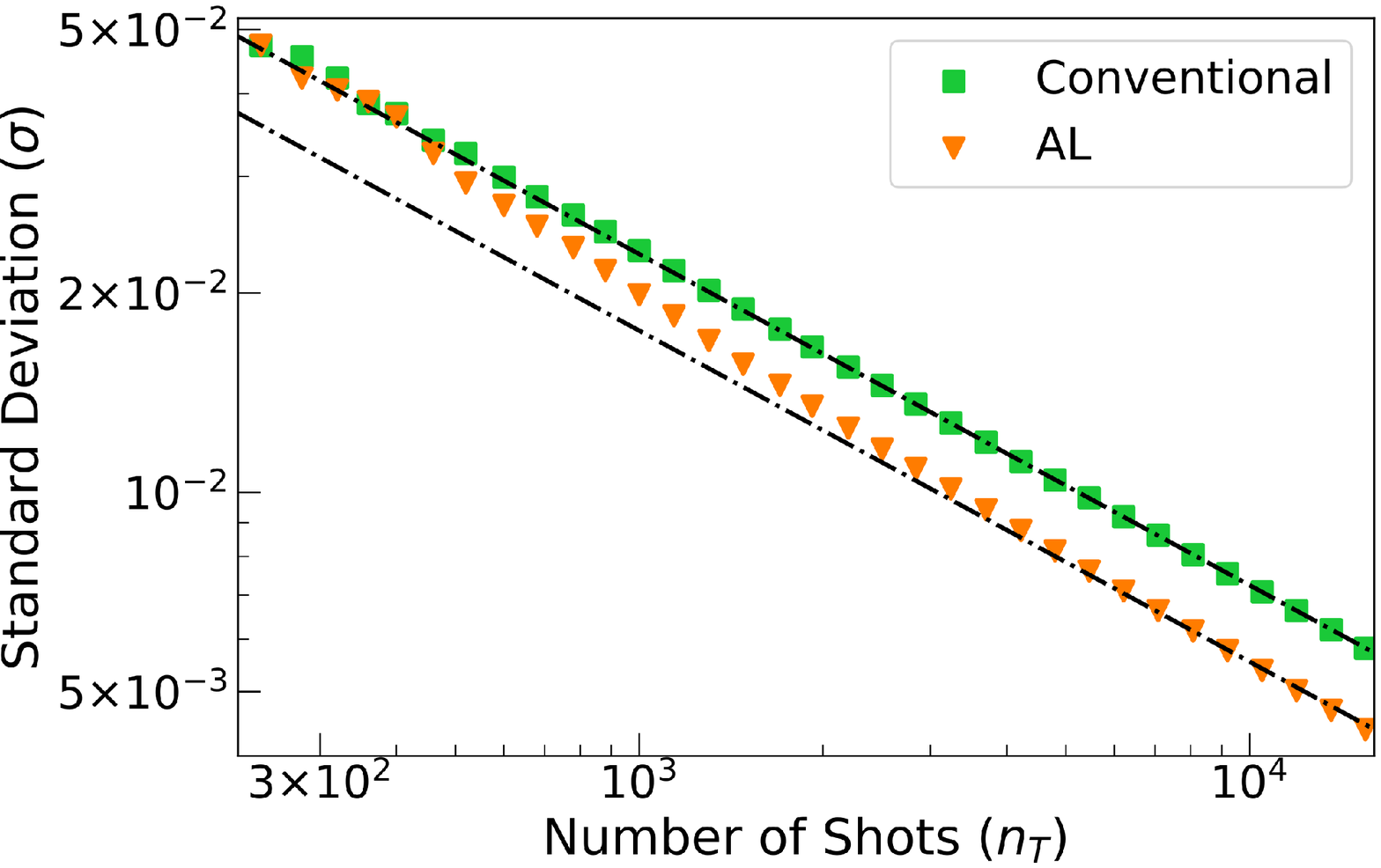}
}
  \hfill
  \subfloat[Toffoli Gate]{
  \includegraphics[width=0.48\textwidth]{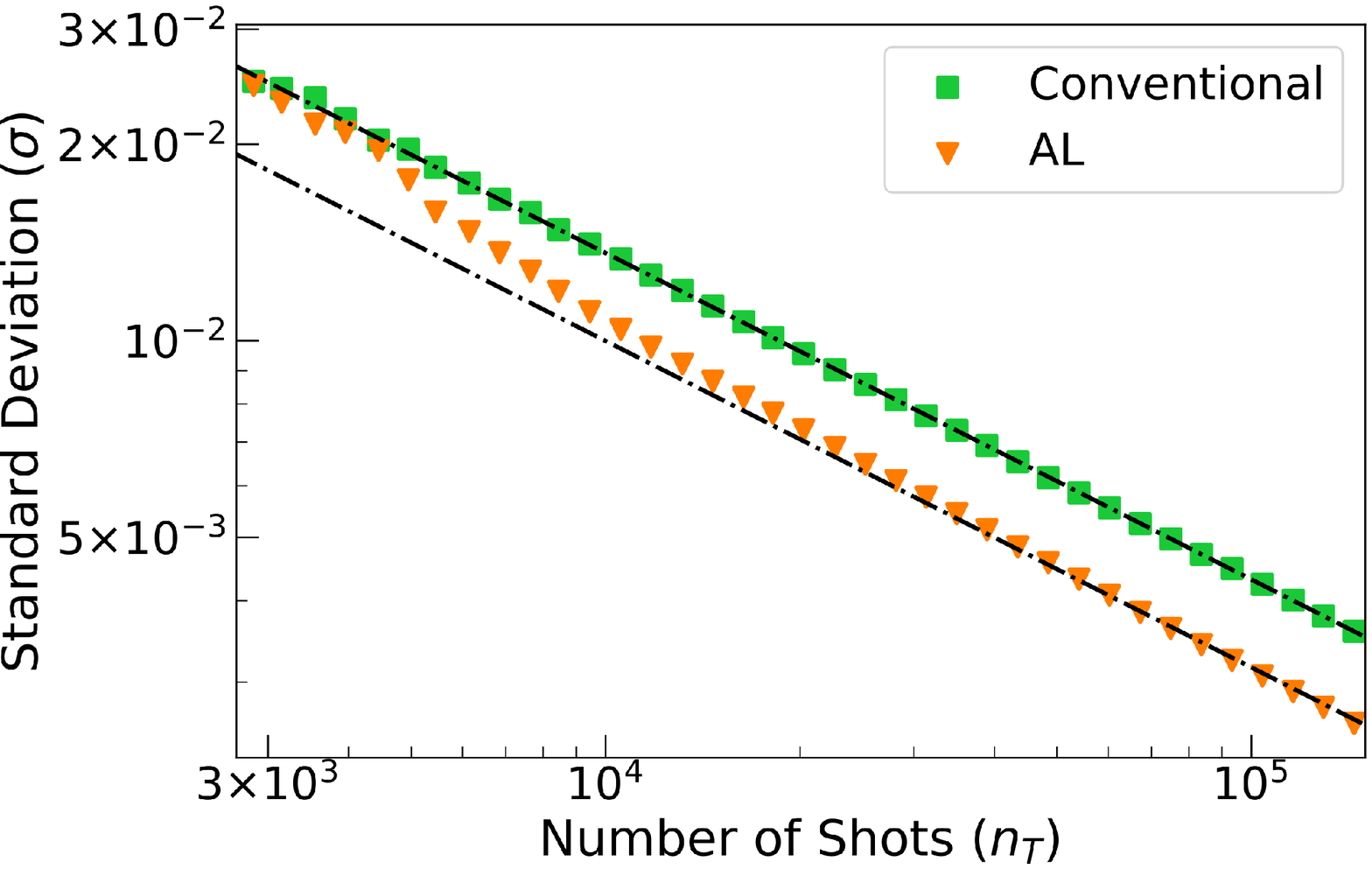}
  }
\caption{
Convergence comparisons of the fidelity estimations with active learning and conventionally uniform allocations of experimental repetitions for (a) CNOT Gate and (b) Toffoli Gate. Both standard deviation and the total number of shots are plotted on a logarithmic scale. The reference lines indicate the convergence that scales as $1/\sqrt{n_T}$.  
}
\label{fig:gate-fidelity}
\end{figure*}

With this choice of states, the gate fidelity for a single qubit can be expressed as
\begin{equation}
    F(\Lambda,U) = \frac{1}{4} \sum_{i,j,k} c_{ijk}
     \bra{i}U^\dagger\Lambda( |\phi_k\rangle \langle \phi_k| )\ U\ket{j} 
     \label{eq:gatefidstates}
\end{equation}
with complex scalar coefficients $c_{ijk}$.
Due to the choice of an over-complete set of states, the values of these coefficients in not uniquely determined, but the choices
$
c_{000}=1, 
c_{111}=1,
c_{01k}=\frac{2}{3}e^{i\frac{2\pi (k-2) }{3}} \mbox{ and } 
c_{10k}=\frac{2}{3}e^{-i\frac{2\pi (k-2) }{3}} \mbox{ for}\ k=2,3,4$,
and all remaining terms vanishing is a valid choice.

This generalises straight-forwardly to the gate fidelity for $N$ qubits, with $5^N$ initial states $\ket{\Phi_k}$ given by tensor products of the single qubits states $\ket{\phi_k}$, and coefficients $C_{ijk}$ given by products of the coefficients $c_{ijk}$.
}

An explicit prescription in terms of state preparation, dynamics described by the channel $\Lambda$ and final measurement is obtained by expanding the operators
$\sum_{ij}C_{ijk} U|j\rangle \langle i|U^\dagger$ into the set of observables that can be directly measured.
With the set of local Pauli measurements $S_i$ also used in Sec.~\ref{sec:statefidelity} for the state fidelity, one obtains
\begin{equation}
    F(\Lambda, U) = \frac{1}{2^{2N}} \sum_{k=0}^{5^n-1} \sum_{l=0}^{4^n-1} a_{lk} \operatorname{tr} \left( S_l \Lambda\left( | \Phi_k \rangle   \langle \Phi_k | \right) \right),
\end{equation}
with
\begin{equation}
    a_{lk} = \frac{ 
    \operatorname{tr} \bigl( S_l \sum_{ij}C_{ijk} U|j\rangle \langle i|U^\dagger \bigr)
    }
    {\tr S_i^2}\ .
\end{equation}

The situation regarding the choice of measurements is thus analogous to the case of state fidelity, but in addition to the choice of measurement, there is also the choice of initial state.
In every step of the process, the AL algorithm will thus select the most informative initial state and corresponding measurement.

Similarly to the estimate of state fidelities discussed in Sec.~\ref{sec:statefidelity}, the accuracy of an empirical estimate of the gate fidelity also depends on the actual realisation of random measurement outcomes,
and a reliable assessment of the two methods to-be-compared is obtained only in terms of statistics of many 
independent realisations of the same fidelity estimates.

Fig.~\ref{fig:gate-fidelity} depicts the decrease of the standard deviation $\sigma(n_T)$ with the number of shots for (a) the estimation of the fidelity between a two-qubit CNOT gate and CNOT channel, and
(b) the estimation of the fidelity between a three-qubit Toffoli gate and a Toffoli channel, similar to Fig.~\ref{fig:convergence}. 
Triangles denote the case of estimates aided by AL, and squares denote the case in which all measurements are taken with the conventional approach. Qualitatively, the convergence confirms the behavior identified in Fig.~\ref{fig:convergence}, but the quantitative details are different:
the period of faster convergence in the approach aided by AL last until $n_T \approx 2000$ for the CNOT gate and until $n_T\approx 1.1\times10^{4}$ for the Toffoli gate.
The improvement as derived from the part of the convergence that satisfies the $1/\sqrt{n_T}$ behaviour is $n_T^{(c)}/n_T^{(AL)}\approx 2$ for the CNOT gate and $n_T^{(c)}/n_T^{(AL)}\approx 2.2$ for the Toffoli gate.
With the larger improvement for the CNOT gate and the Toffoli gate as compared to the improvement found for two-qubit and three-qubit state fidelities,
Fig.~\ref{fig:gate-fidelity} thus confirms the expectations that the benefits of AL are growing with the number of measurement settings to choose from.

\section{Outlook}

In particular, in the era of {\em noisy intermediate-scale quantum} (NISQ) devices~\cite{Preskill2018NISQ}, the estimate of state-fidelities and gate fidelities is a commonly encountered problem\cite{Zhu2022NoiseState,Zhang2020NoiseGate,Noiri2022NoiseGate}.
The rapidly growing number of observables to be measured makes this an extremely challenging task even for moderate qubit numbers\cite{Lu2015Difficulty,Zhou2020Difficulty}.
Due to the large noise level in such devices\cite{MARTINA2022Noise,Almudever2017Noise}, there is large uncertainty about a created state or an implemented gate, so a prior allocation of measurement repetitions for the specific state or gate is indeed problematic. The interactive active learning process for observables to be measured can thus practically facilitate the estimate of fidelities.
Since such estimates are at the core of data-driven optimisation processes~\cite{Dive2018insituupgradeof,PRXQuantum.1.020322,Omran2019}, and due to their iterative nature, these optimisations require several fidelity estimates,
the proposed algorithm can contribute to our ability to derive practical use from faulty hardware.

The use of the proposed techniques is also not limited to fidelities, but it can also find applicability in {\em variational quantum algorithms} (VQA)~\cite{Cerezo2021VQA,Wecker2015VQA,Kandala2017VQA} in which expectation values of a Hamiltonian or some other operator need to be estimated.
The goal of the VQA is the experimental realisation of the quantum state that minimises this expectation value,
and iterative optimisation algorithms estimate this expectation value with the same accuracy for all considered states~\cite{Self2021}.
Since the proposed algorithm does provide not only empirical expectation values, but also bounds on their accuracy,
it can also identify the lowest conceivable expectation value at any point in time during the data acquisition.
As soon as this value exceeds the expectation value observed with another state, one can safely stop taking data based on this state and start estimating the expectation value with a different state.

With possible extensions to the estimate of quantities like entropy or correlation functions involving products of expectation values,
the use of the proposed active learning algorithm has clear potential to become a commonly used tool in the analysis of quantum systems.

\section*{Acknowledgements} 
We are indebted to stimulating discussions with Zezhen Wei. Numerical simulations were carried out on Imperial College High-Performance Computing facilities~\cite{HPC}. 

\bibliography{references}

\end{document}